\documentclass[12pt]{article}
\usepackage{ametsoc}
\usepackage{wasysym}
\usepackage{textcomp}
\usepackage{amssymb}
\usepackage{graphicx}
\usepackage{lineno}

\usepackage{color,soul}
\definecolor{lightblue}{rgb}{.80,.85,1}
\sethlcolor{lightblue}

\linenumbers
\newcommand{\myabstract}
{Studies over a period of several decades have resulted in a relatively simple set of equations describing the tidally and width-averaged balances of momentum and salt in a rectangular estuary. We rewrite these equations
 in a fully non-dimensional form that yields two non-dimensional variables: (i) the estuarine Froude number; and
(ii) a modified tidal Froude number.  The latter is the product of the tidal Froude number and the square root of the estuarine aspect ratio.  These two variables are used to define a prognostic estuary classification scheme, which compares
favourably with published estuarine data.}

\begin{document}
%

\title{\textbf{\large{Estuary Classification Revisited}}}
%
%
\author{\textsc{Anirban Guha}
				\thanks{\textit{Corresponding author address:} 
				Anirban Guha, Civil Engineering Department, The University of British Columbia,
				Vancouver, B.C., Canada V6T 1Z4. 
				\newline{E-mail:aguha@mail.ubc.ca}}
\quad\textsc{and Gregory A. Lawrence}\\
\textit{\footnotesize{Civil Engineering Department, The University of British Columbia, Vancouver, B.C., Canada}}\\
\textit{\footnotesize{\& Institute of Applied Mathematics, The University of British Columbia, Vancouver, B.C., Canada}}
}
%
\ifthenelse{\boolean{dc}}
{
\twocolumn[
\begin{@twocolumnfalse}
\amstitle

\begin{center}
\begin{minipage}{13.0cm}
\begin{abstract}
	\myabstract
	\newline
	\begin{center}
		\rule{38mm}{0.2mm}
	\end{center}
\end{abstract}
\end{minipage}
\end{center}
\end{@twocolumnfalse}
]
}
{
\amstitle
\begin{abstract}
\myabstract
\end{abstract}
\newpage
}
\section{Introduction}

Since the introduction of  stratification-circulation diagram by  \cite{HR1966}, numerous estuarine classification schemes have been proposed.  The reader might ask - why revisit this topic?  Our motivation for
pursuing a new classification scheme stems from notable recent advances in estuarine physics, many of which are reviewed in  \cite{MG2010}.  These advances led us to hypothesize that there might be a simple
means to determine the conditions under which a sufficiently well behaved estuary will be well mixed, partially mixed, or highly stratified.
We start by outlining the classical tidally averaged model as presented by  \cite{MG2010}.  We then rewrite the equations of this model in non-dimensional form.  Using this new set of equations we develop our classification scheme,
and then compare its predictions with field observations.

\section{Classical Tidally Averaged Model}
\label{sec:model}

The physics of estuarine circulation is governed by the  competing influences of river and oceanic flows. While the former adds fresh water, the latter adds denser salt water which moves landward due to  the combined effect of tides and gravitational circulation (or exchange flow).   The complicated balance between the river, the exchange flow and the tides determines the estuarine velocity and salinity structure. 

We consider an idealized rectangular estuary of depth $H$ and width $B$. The origin of the coordinate system is at the free surface at the mouth of the estuary with 
 the horizontal ($x$) axis pointing seawards and the vertical ($z$) axis pointing upwards.
Therefore, both the horizontal and vertical distances within the estuary  are negative quantities. 
To obtain the width-averaged and tidally-averaged horizontal
velocity ($u$), and salinity ($s$) distribution in the estuary, these quantities are first decomposed into depth averaged (overbar) and  depth varying (prime) components: $u=\bar{u}(x,t)+u'(x,z,t)$,
$s=\bar{s}(x,t)+s'(x,z,t)$. The quantity $\bar{u}= Q_{R}/A$ is the cross-sectionally averaged river velocity, where $Q_{R}$ is the mean river flow rate and $A= B H$. The solution for both  partial and well mixed estuaries was given by
\cite{HR1965} (for recent review, see \cite{MG2010}): 
\begin{linenomath*}
\begin{eqnarray}
& & u=\bar{u}+u'=\bar{u}P_{1}+u_{E}P_{2}\label{eq:u}\\
& & s=\bar{s}+s'=\bar{s}+\frac{H^{2}}{K_{S}}\bar{s}_{x}\left(\bar{u}P_{3}+u_{E}P_{4}\right) \label{eq:s}
\end{eqnarray}
\end{linenomath*}
\begin{linenomath*}
\begin{eqnarray}
\textrm{where} & &  P_{1}=\frac{3}{2}-\frac{3}{2}\xi^{2}  \nonumber \\ 
& & P_{2}=1-9\xi^{2}-8\xi^{3}\nonumber \\
& & P_{3}=-\frac{7}{120}+\frac{1}{4}\xi^{2}-\frac{1}{8}\xi^{4}\nonumber \\
& & P_{4}=-\frac{1}{12}+\frac{1}{2}\xi^{2}-\frac{3}{4}\xi^{4}-\frac{2}{5}\xi^{5}
\label{eq:xii}
\end{eqnarray}
\end{linenomath*}
 In (\ref{eq:xii}), $\xi = z/H \in\left[-1,0\right]$ is the normalized vertical coordinate. The subscript $x$ implies $\partial/\partial x$ where $x$ is dimensional.
 $K_{S}$ is the vertical eddy diffusivity.
 For exchange dominated estuaries, an important parameter is the exchange velocity scale:
 \begin{linenomath*}
\begin{equation}
u_{E} = c^{2}H^{2}\bar{\Sigma}_{x}/\left(48K_{M}\right)
\label{eq:ue}
\end{equation}
\end{linenomath*}
Here $c=\sqrt{g\beta s_{ocn}H}$  is twice the speed of the fastest internal wave that can be supported in an estuary \cite[]{MG2010}.  $K_{M}$ is the vertical eddy 
viscosity and $\beta \cong 7.7\times10^{\text{-}4}$ psu$^{-1}$. The non-dimensional salinity is defined 
as $\Sigma = s/s_{ocn}$, 
where $s_{ocn}$ is the ocean salinity.

Equations (\ref{eq:u})-(\ref{eq:s}) were derived under the assumption that the density field is governed by the linear equation of state: 
$\rho=\rho_{0}\left(1+\beta s\right)$ where $\rho_{0}$ is the density of fresh water. The details of the derivation 
are  well documented in \cite{M1999,M2004}.

The salt balance is given by:
\begin{linenomath*}
\begin{eqnarray}
\underbrace{\frac{d}{dt}\int\Sigma\, dx}\,\,\,\, & = & \,\,-\underbrace{\overline{u'\Sigma'}}\,\,\,\,\,+\,\,\,\underbrace{K_{H}\bar{\Sigma}_{x}}\,\,\,\,\underbrace{-\bar{u}\bar{\Sigma}}  \nonumber \\
\mathsf{{\normalcolor \textrm{accumulation}}} & \, & \,\textrm{{\normalcolor exchange\,\,\,\,\,\,\,\ \, tidal}\,\,\,\,\,\, river} 
\label{eq:10}
\end{eqnarray}
\end{linenomath*}
where, $K_{H}$ is the horizontal diffusivity. This equation physically implies  that the temporal salt accumulation in an estuary is due to the competition between salt addition and removal processes.
While exchange (note that $\overline{u'\Sigma'}$ 
is negative) and tidal processes add salt, 
river inflow removes it. At steady state (\ref{eq:10})  can be rewritten as:
\begin{linenomath*}
\begin{eqnarray}
\underbrace{\bar{\Sigma}}=\underbrace{\left(L_{E3}\bar{\Sigma}_{x}\right)^{3}}+\,\underbrace{\left(L_{E2}\bar{\Sigma}_{x}\right)^{2}}+\underbrace{L_{E1}\bar{\Sigma}_{x}}+\underbrace{L_{H}\bar{\Sigma}_{x}}\nonumber \\
\textrm{R\,\,\,\,\,\,\,\,\,\,\,\,\,\,\,\ \,\,\,\,\ensuremath{E_{3}}\ \,\,\,\,\,\,\,\,\ \,\,\,\ \,\,\,\ \,\,\ensuremath{E_{2}}\,\,\,\,\,\,\,\,\,\,\ \,\,\,\ \,\,\ \ensuremath{E_{1}}\ \,\,\ \,\,\,\,\,\,\,\,\,\,\ T\,\,\ }\textrm{ }
\label{eq:11}
\end{eqnarray}
\end{linenomath*}
\begin{linenomath*}
\begin{eqnarray}
\textrm{where}  &  & L_{H}=K_{H}/\bar{u} \nonumber \\
 &  & L_{E1}=0.019\bar{u}H^{2}/K_{S} \nonumber \\
 &  & L_{E2}=0.031cH^{2}/\left(K_{S}K_{M}\right)^{1/2} \nonumber \\
 &  & L_{E3}=0.024(c/\bar{u})^{1/3}cH^{2}/\left(K_{S}K_{M}^{2}\right)^{1/3}
\label{eq:12}
\end{eqnarray}
\end{linenomath*}
The different terms in  (\ref{eq:11}) are as follows:  \ensuremath{R} is the river term, \ensuremath{T} is the tidal term, while \ensuremath{E_{1}}, \ensuremath{E_{2}} and \ensuremath{E_{3}} are the different components of the exchange term.
 \cite{HR1965}  presented (\ref{eq:11}) in a slightly different form, and \cite{M2004,M2007} introduced the length scales in (\ref{eq:12}).

The length scales in (\ref{eq:12}) depend upon the mixing co-efficients: $K_{S}$, $K_{M}$ and $K_{H}$.  Making use of an extensive study of  Willapa Bay, \cite{Banas2004} proposed:
 \begin{linenomath*}
\begin{equation}
K_{H}=a_{1}u_{T}B;
\label{eq:kt}
\end{equation}
\end{linenomath*}
where $a_{1}=0.035$ and $u_{T}$ is the amplitude of the depth averaged tidal flow. Based on field studies and modeling of the Hudson River estuary,~\cite{Ralston2008} obtained
\begin{linenomath*}
\begin{equation}
K_{M}=a_{0}C_{D}u_{T}H\,\mathrm{and}\, K_{S}=K_{M}/Sc;
\label{eq:km}
\end{equation}
\end{linenomath*}
where $a_{0}=0.028$,~$C_{D}=0.0026$ and $Sc=2.2$ is a Schmidt number. We will use (\ref{eq:kt}) and (\ref{eq:km}) in the development of a non-dimensional set of equations.

While the governing equations (\ref{eq:u}), (\ref{eq:s}) and (\ref{eq:11})  are elegant representations of the problem of estuarine circulation, they are sufficiently complicated that simplifications have been sought after.  Numerous investigators, including  \cite{HR1965,Chat1976,Moni2002,M2004,MG2010} have assumed $\bar{u}\ll u_{E}$, which yields:
\begin{linenomath*}
\begin{eqnarray}
\underbrace{\bar{\Sigma}}=\underbrace{\left(L_{E3}\bar{\Sigma}_{x}\right)^{3}}+\underbrace{L_{H}\bar{\Sigma}_{x}}  \nonumber\\
\textrm{ R\,\textrm{\,\,\ \ \,\,\,\,\,\ \,\ \,\ \,\ensuremath{E_{3}}\ \,\,\,\ \ \,\,\,\,\ \,\,\ \ T\,\,\ \,\ }}
\label{eq:choto}
\end{eqnarray}
\end{linenomath*}
\cite{Chat1976} further reduced (\ref{eq:choto}) to two simple cases with analytical solutions, the exchange dominated case ($\textrm{T}\rightarrow0$), and the tidally dominated case ($\textrm{\ensuremath{E_{3}}}\rightarrow0$).  While these approximations have been widely used there does not appear to have been any serious attempt to determine the conditions under which they are applicable.

\section{Non-dimensional Tidally Averaged Model}
\label{sec:salt}

In this section we rewrite the governing equations (\ref{eq:u}), (\ref{eq:s}) and (\ref{eq:11})  in non-dimensional form in anticipation of (i) revealing the important non-dimensional parameters governing the problem, and (ii) facilitating comparison of the relative magnitude of each of the terms in (\ref{eq:11}).  Defining $X = x/L_{E3}$, (\ref{eq:11}) can be rewritten as:
\begin{linenomath*}
\begin{equation}
\bar{\Sigma}=\bar{\Sigma}_{X}^{3}+\left(\frac{L_{E2}}{L_{E3}}\right)^{2}\bar{\Sigma}_{X}^{2}+\left(\frac{L_{E1}}{L_{E3}}\right)\bar{\Sigma}_{X}+\left(\frac{L_{H}}{L_{E3}}\right)\bar{\Sigma}_{X}
\label{eq:intermed}
\end{equation}
\end{linenomath*}
\begin{linenomath*}
\begin{eqnarray}
 \textrm{where} &  & \left(\frac{L_{E2}}{L_{E3}}\right)^{2}=\left(\frac{0.031}{0.024}\right)^2Sc^{1/3}F_{R}^{2/3}=2.17F_{R}^{2/3} \nonumber \\
 &  & \frac{L_{E1}}{L_{E3}}=\left(\frac{0.019}{0.024}\right)Sc^{2/3}F_{R}^{4/3}=1.34F_{R}^{4/3}  \nonumber \\
 &  & \frac{L_{H}}{L_{E3}}=\left(\frac{a_{0}a_{1}C_{D}}{0.024}\right)Sc^{-1/3}\left(B/H\right)F_{T}^{2}F_{R}^{-2/3} 
\label{eq:ratios}
\end{eqnarray}
\end{linenomath*}
The velocity $\bar{u}$ and $u_{T}$ have been non-dimensionalized by $c$ to obtain the densimetric estuarine Froude number $F_{R}=\bar{u}/c$ 
and the tidal Froude number $F_{T}= u_{T}/c$. Substituting (\ref{eq:ratios}) into (\ref{eq:intermed}) yields:
\begin{linenomath*}
\begin{eqnarray}
\underbrace{\bar{\Sigma}}=\underbrace{\bar{\Sigma}_{X}^{3}}+\,\underbrace{ C_{1}F_{R}^{2/3}\bar{\Sigma}_{X}^{2}}+\underbrace{ C_{2}F_{R}^{4/3}\bar{\Sigma}_{X}}+\underbrace{C_{3}\widetilde{F_{T}}^{2}F_{R}^{-2/3}\bar{\Sigma}_{X}}\nonumber \\
\textrm{R}\,\,\,\,\,\,\,\,\,\,\,\,\,\,\ensuremath{E_{3}}\,\,\,\,\,\,\,\,\,\,\,\,\,\,\,\,\,\,\,\ensuremath{E_{2}}\,\,\,\,\,\,\,\,\,\,\,\,\,\,\,\,\,\,\,\,\,\,\,\,\,\,\ensuremath{E_{1}}\,\,\,\,\,\,\,\,\,\,\,\,\,\,\,\,\,\,\,\,\,\,\,\ \ \,\,\ \,\,\, \textrm{T }\,\,\,\ \,\,\,\,\,\,\,\,\,\,\,
\label{eq:ashol}
\end{eqnarray}
\end{linenomath*}
where  $C_{1}=2.17$, $C_{2}=1.34$, $C_{3}=8.16\times10^{-5}$, and the \emph{modified} tidal Froude number, $\widetilde{F_{T}}= F_{T} \sqrt{B/H}$. 
Typically the estuarine aspect ratio $B/H\sim O\left(10^{2}-10^{3}\right)$, see Table \ref{tab:t1}. The  magnitude of different terms in (\ref{eq:ashol}) can be easily compared by noting that $0<O\left(Fr^{4/3}\right)<O\left(Fr^{2/3}\right)<O\left(1\right)<O\left(Fr^{-2/3}\right)$.  The tidal term (T)  however depends on an additional parameter $\widetilde{F_{T}}$, whose (order of) magnitude needs to be known for making the comparison. 

Like the salt balance equation, the momentum  and salinity equations, i.e.\  (\ref{eq:u}) and (\ref{eq:s}) can also be expressed in non-dimensional form as follows: 
\begin{linenomath*}
\begin{align}
 U=C_{4}F_{R}^{1/3}\bar{\Sigma}_{X}\, P_{2}+F_{R}P_{1} \label{eq:newu1} \,\ \,\,\,\,\,\,\,\,\,\,\ \,\,\,\,\,\,\,\,\,\,\ \,\,\,\, \\
\Sigma=\bar{\Sigma}+C_{5}F_{R}^{2/3}\bar{\Sigma}_{X}^{2}\, P_{4}+C_{6}F_{R}^{4/3}\bar{\Sigma}_{X}\,P_{3} \label{eq:newsigma1}
\end{align}
\end{linenomath*}
The constants $C_{4}=0.667$, $C_{5}=47.0$ and $C_{6}=70.5$. In (\ref{eq:newu1}), the quantity $ U = u/c$ is the non-dimensional horizontal velocity (not to be confused with $F_{R}$, which is $\bar{u}/c$). 
Equations (\ref{eq:ashol})-(\ref{eq:newsigma1}) are the non-dimensional governing equations for our idealized estuary.

\begin{figure}[t]
  \centering
\includegraphics[trim=0cm 0cm 0cm 0cm, clip=true,scale=0.42]{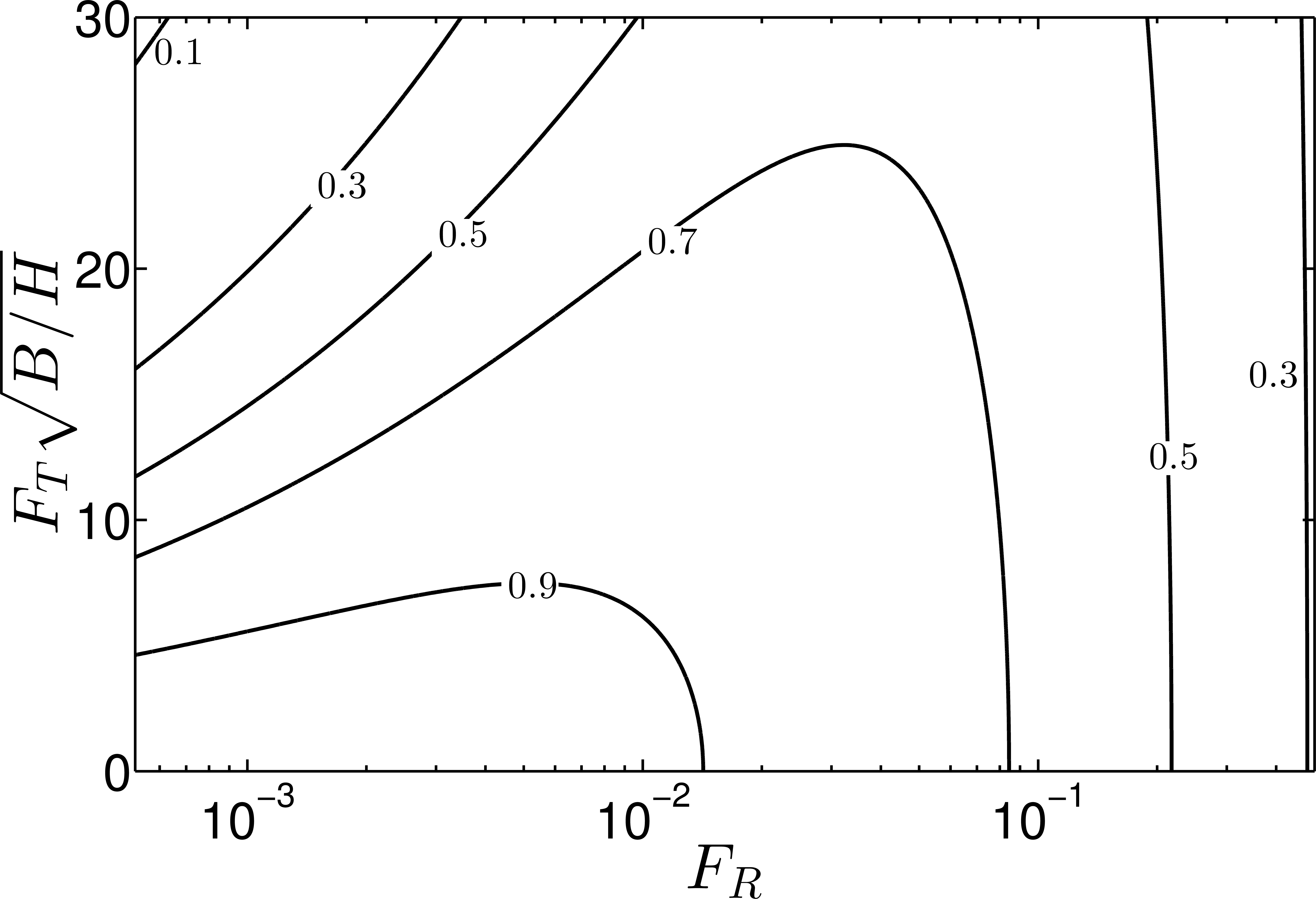}
 \caption{The variation of salinity gradient at the estuary mouth ($\bar{\Sigma}_{X}|_{0}$)  with the estuary control variables - $F_{R}$ and $\widetilde{F_{T}}$. The solid lines represent isocontours of  $\bar{\Sigma}_{X}|_{0}$. 
}\label{fig:Sigma_X0}
\end{figure}

Eq.\ (\ref{eq:ashol}) poses a non-linear initial value problem which can only be solved numerically. For that, the  conditions at the estuary mouth have to be determined. One such condition is $\Sigma\left(0,-1\right)=1$; 
meaning the salinity at the bed of the estuary at its mouth has to be the same as the ocean salinity. Substituting (\ref{eq:newsigma1}) into  (\ref{eq:ashol})  and making use of this condition, we obtain
\begin{linenomath*}
\begin{equation}
\left(\bar{\Sigma}_{X}|_{0}\right)^{3}+C_{7}F_{R}^{2/3}\left(\bar{\Sigma}_{X}|_{0}\right)^{2}+\left(C_{8}F_{R}^{4/3}+C_{3}\widetilde{F_{T}}^{2}F_{R}^{-2/3}\right)\bar{\Sigma}_{X}|_{0}=1;
\label{eq:Full1}
\end{equation}
\end{linenomath*}
where $C_{7}=5.31$ and $C_{8}=6.04$. Eq.\ (\ref{eq:Full1}) is actually the non-dimensional version of Eq. (19) of \cite{M2004}. Being a cubic equation, it can be solved analytically to evaluate the salinity gradient at the estuary mouth, $\bar{\Sigma}_{X}|_{0}$. 
Additionally, (\ref{eq:Full1}) indicates that $\bar{\Sigma}_{X}|_{0}$ is only a function of $F_{R}$ and $\widetilde{F_{T}}$. The variation of $\bar{\Sigma}_{X}|_{0}$ with
these two Froude numbers is depicted in Fig.\ \ref{fig:Sigma_X0}. The figure shows that $0<\bar{\Sigma}_{X}|_{0}<1$ over the entire parameter space.

\section{Estuary Classification} 
\label{sec:classify}

\begin{figure}[t]
  \centering
\includegraphics[trim=0cm 0cm 0cm 0cm, clip=true,scale=0.3]{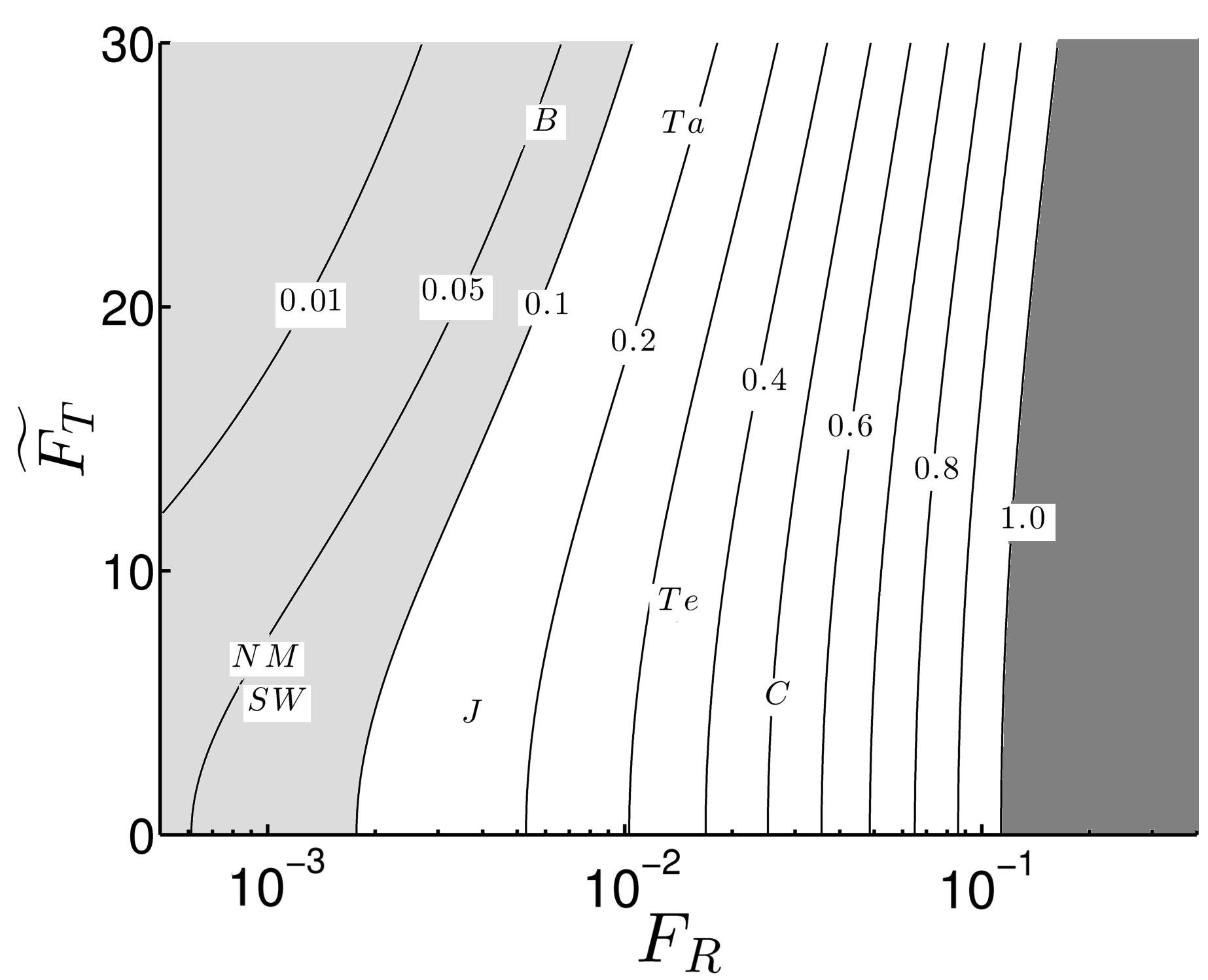}
 \caption{Estuary Classification Diagram. The lines represent isocontours of  $\Phi_{0}$.  The three regions represent three types of estuaries: (a) light Grey- well mixed, (b) White - partially mixed, and (c) dark Grey- highly stratified or salt wedge. The letters denote estuaries: 
  C - Columbia, J - James, Te - Tees, SW - Southampton Waterway, Ta - Tay, 
NM - Narrows of the Mersey and B - Bristol Channel. For data, see Table \ref{tab:t1}.
}\label{fig:classification2}
\end{figure}

Our goal is to develop a simple  classification scheme that distinguishes between well-mixed, partially mixed and highly stratified estuaries. A relevant  parameter for classifying estuaries is the 
non-dimensional salinity stratification at the estuary mouth, $\Phi_{0}$. It is defined as follows:
\begin{linenomath*}
\begin{equation}
\Phi_{0} = \Sigma\left(0,-1\right)-\Sigma\left(0,0\right)
\label{eq:phi0}
\end{equation}
\end{linenomath*}
This parameter ranges between $0$ and $1$. While the lower limit implies a very well mixed estuary,  the upper limit indicates the transition to salt wedge. 
Substituting (\ref{eq:newsigma1}) into (\ref{eq:phi0}) yields:
\begin{linenomath*}
\begin{equation}
\Phi_{0}=C_{9}F_{R}^{2/3}\left(\bar{\Sigma}_{X}|_{0}\right)^{2}+C_{10}F_{R}^{4/3}\bar{\Sigma}_{X}|_{0};
\label{eq:Full2}
\end{equation}
\end{linenomath*}
where $C_{9}=7.06$ and $C_{10}=8.82$. If $F_{R}$ and  $\widetilde{F_{T}}$ are known, then $\bar{\Sigma}_{X}|_{0}$ can be directly obtained by
solving (\ref{eq:Full1}).
Consequently, $\Phi_{0}$ can be evaluated from (\ref{eq:Full2}), yielding Fig.\ \ref{fig:classification2}.

We follow \cite{HR1966} and use the condition $\Phi_{0}=0.1$ to define the transition between well mixed and partially mixed estuaries.
To distinguish between partially mixed and highly stratified estuaries we use the condition $\Phi_{0}=1.0$, 
corresponding to fresh surface water extending to the mouth of the estuary.  Our classification scheme is obtained by plotting these transitional criteria on Fig.\ \ref{fig:classification2}. 
When $\widetilde{F_{T}}=0$, the transition between well-mixed and partially-mixed estuaries is predicted to occur at $F_{R}=0.0017$, and from partially-mixed to highly stratified at $F_{R}=0.113$. 
The value of $F_{R}$ for both transitions increases as $\widetilde{F_{T}}$ increases, the increase being more rapid for the transition from partially-mixed to highly stratified estuaries.  These results are in qualitative 
agreement with Fig. 2.7 of \cite{Estuarybook}. 


\section{Discussion} 
\label{sec:disc}

Together  (\ref{eq:Full1}) and (\ref{eq:Full2}) provide new insight into estuarine physics. Apart from broadly classifying estuaries into three categories, viz.\ highly stratified, partially mixed and well mixed,  
the equation set  identifies $F_{R}$ and $\widetilde{F_{T}}$ to be the \emph{only} two parameters determining the  stratification at the  estuary mouth, $\Phi_{0}$. The new non-dimensional parameter  $\widetilde{F_{T}}= F_{T} \sqrt{B/H}$  reveals  that ``tidal effect'' is not simply represented by the tidal Froude number $F_{T}$, but the latter combined with the square-root of the estuarine aspect ratio $B/H$. Moreover the equation set  \emph{predicts} $\Phi_{0}$, given $F_{R}$ and $\widetilde{F_{T}}$. If estuarine condition changes, e.g.\  river flow changes from low to high, tidal flow changes from spring to neap, or estuary depth changes due to dredging, the parameters $F_{R}$ and $\widetilde{F_{T}}$ will change correspondingly. These newly obtained Froude numbers will produce  a new $\Phi_{0}$, which reflects the response of estuarine circulation and mixing to variability.

\begin{figure}[t]
  \centering
\includegraphics[trim=0.0cm 0cm 0cm 0cm, clip=true,scale=0.30]{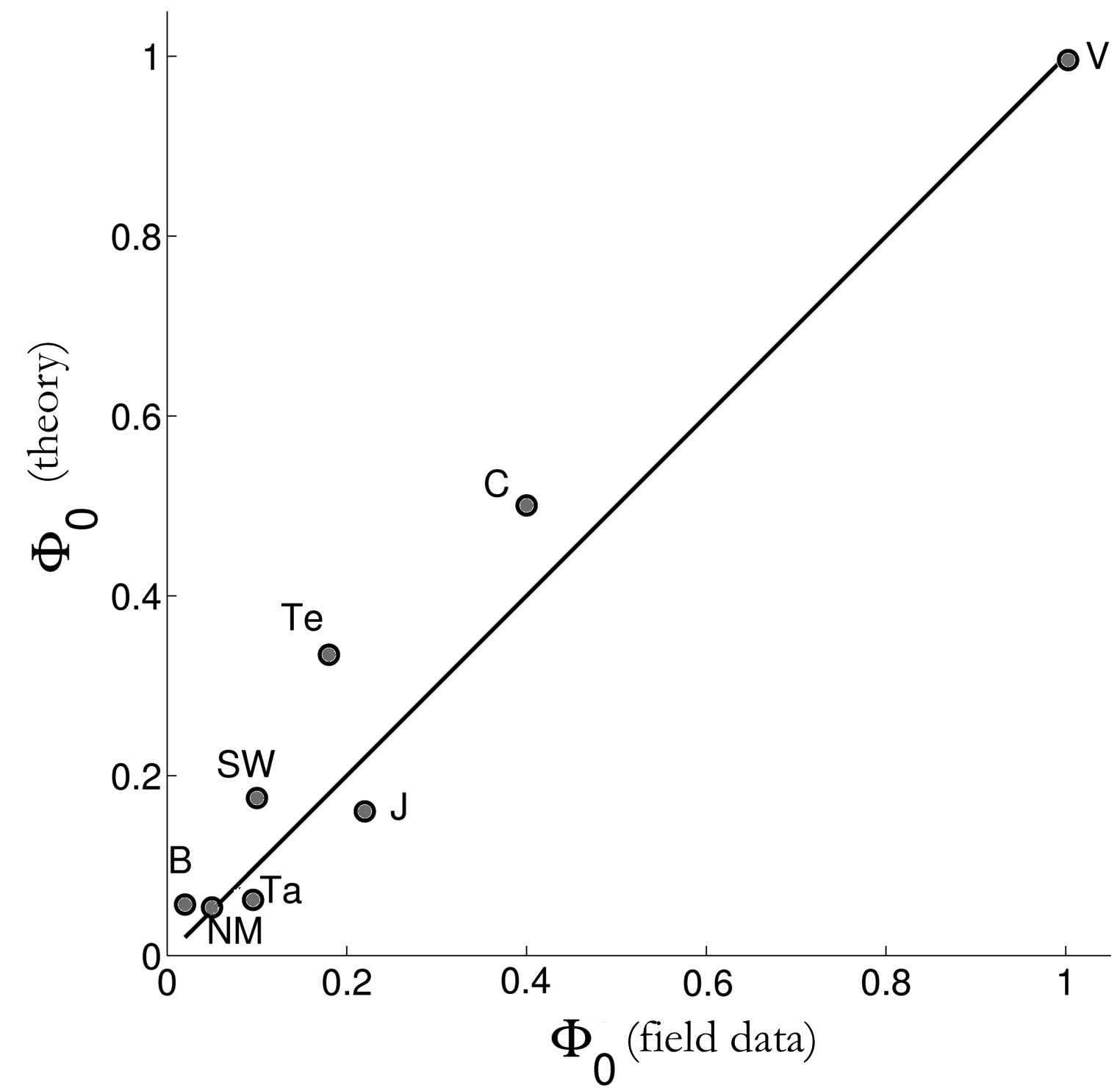}
 \caption{Comparison between stratification at the estuary mouth obtained from theory with field data).
}\label{fig:last}
\end{figure}

To test the applicability of our classification scheme we made use of the field data presented in \cite{Pran85}. 
Using these data we have computed $F_{R}$, $F_{T}$, $B/H$, and $\widetilde{F_{T}}$ directly, and $\Phi_{0}$ from (\ref{eq:Full1}) and (\ref{eq:Full2}); see Table \ref{tab:t1}. We have compared the 
computed value of $\Phi_{0}$  with the measured value in Fig. \ref{fig:last}. 
The comparison is good considering 
the accuracy to which $F_{R}$  and $\widetilde{F_{T}}$ can be determined from field data.

\begin{figure}[t]
 \centering
\includegraphics[trim=0.0cm 0cm 0cm 0cm, clip=true,scale=0.22]{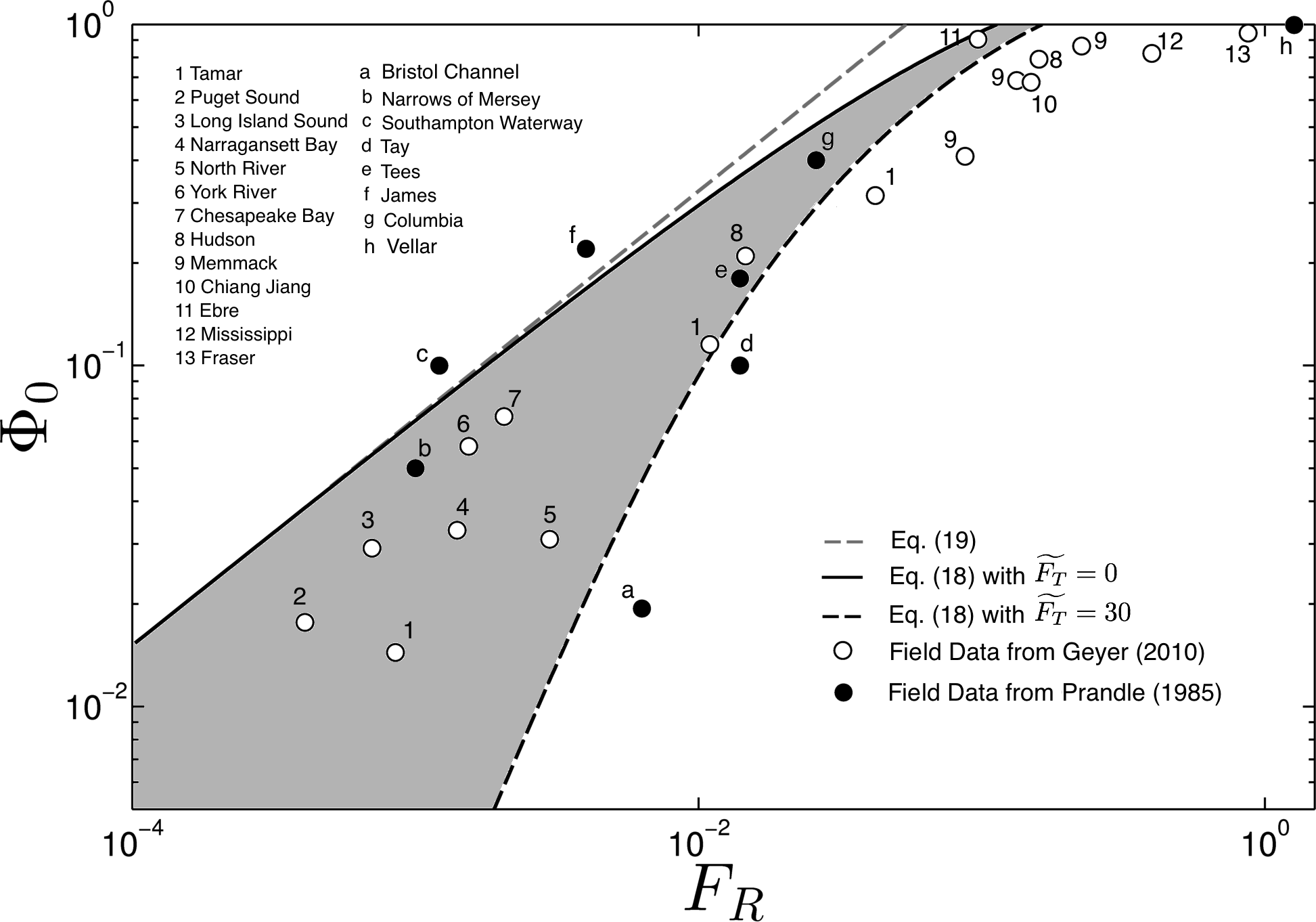}
 \caption{Comparison between our estuary classification scheme and the approximation $\Phi_{0}=7F_{R}^{2/3}$ in (\ref{eq:stateofart}). The grey area indicates the region where estuaries should ideally cluster. 
Field data from \cite{Estuarybook} and \cite{Pran85} are plotted for comparison with the theoretical predictions.   
}\label{fig:comparison}
\end{figure}

It is interesting to note that if both $F_{R}$  and $\widetilde{F_{T}}$  are small then (\ref{eq:Full1}) reduces to $\bar{\Sigma}_{X}|_{0}=1$ and (\ref{eq:Full2}) reduces to:
\begin{linenomath*}
\begin{equation}
\Phi_{0}\approx7F_{R}^{2/3};
\label{eq:stateofart}
\end{equation}
\end{linenomath*}
which is exactly the same as Eq. (19) of \cite{MG2010}.  They have found this equation by combining Knudsen's relations \cite[]{Knud00} with (\ref{eq:km}). Since Knudsen's relations are derived from mass and salt balances and do not consider momentum balance, (\ref{eq:stateofart}) provides a rather simplistic prediction. 
We compare (\ref{eq:stateofart}) with the exact solution of (\ref{eq:Full2}) for
both  $\widetilde{F_{T}}=0$ and $\widetilde{F_{T}}=30$ in Fig.\ \ref{fig:comparison}.  When  $\widetilde{F_{T}}=0$,   (\ref{eq:stateofart}) is very accurate up to the 
transition between well-mixed and partially mixed estuaries ($\Phi_{0}=0.1$).  The deviation between 
(\ref{eq:stateofart}) and the exact solution increases with increasing $F_{R}$. Eq.\ (\ref{eq:stateofart}) predicts the transition between partially mixed and highly stratified estuaries 
($\Phi_{0}=1.0$) at $F_{R}=0.054$ rather than at $F_{R}=0.113$.  When $\widetilde{F_{T}}=30$, 
(\ref{eq:stateofart})  is always less than the exact solution in terms of $\Phi_{0}$.

 We also compare the theoretical results with the field data of \cite{Pran85} in Table \ref{tab:t1} and Fig.\ 2.6 of  \cite{Estuarybook} in Fig.\ \ref{fig:comparison}.  This comparison is  mainly intended to provide a qualitative estimate. We have chosen to plot (\ref{eq:Full2}) for $\widetilde{F_{T}}=0$ and $30$, since estuaries mostly have $\widetilde{F_{T}}$ within this range. Ideally, 
most of the partially and well mixed estuaries should cluster within the grey region bounded by the lines $\widetilde{F_{T}}=0$ and $30$, which is indeed the case. The most important aspect of this comparison is that the theoretical curves follow the  overall trend of the field data. However these curves  grossly over-predict $\Phi_{0}$,  therefore they under-predict vertical mixing. This discrepancy may arise if the values of $\Phi_{0}$ were measured at an upstream location, rather than at the mouth (which might be the case for the data points of \cite{Estuarybook}). 
Surprisingly, disagreement between theory and field data does not appear in Fig.\ 2.6 of \cite{Estuarybook}. The latter figure shows that the line, referred to as ``Eq.\ (2.22)'', matches very well with the data points. Although Eq.\ (2.22) is actually
$\Phi_{0}=8.73F_{R}^{2/3}$ (we calculated the coefficient from the associated text in \cite{Estuarybook}), it is mistakenly plotted as $\Phi_{0}\approx 3F_{R}^{2/3}$.

Finally we refer to the assumptions behind our theoretical analyses and their consequences. We have simplified the problem by assuming a tidally averaged estuary with rectangular geometry. In real estuaries  bathymetry can play a crucial role in determining the estuarine circulation.  
Moreover the appearance of just two parameters ($F_{R}$ and $\widetilde{F_{T}}$) in our equations is a consequence of the empirical  equations 
 (\ref{eq:kt}) and (\ref{eq:km}). These two equations are also used in determining the coefficients  $C_{1}, C_{2},\ldots, C_{10}$. All these coefficients are found to depend upon $Sc$, making it the most important parameter in this regard; see Table \ref{tab:t2}. Following \cite{Ralston2008}, $Sc=2.2$ in all our calculations. 
Although  (\ref{eq:kt}) and (\ref{eq:km}) are simple and elegant, they may not be very realistic. In real estuaries both $K_{M}$ and $K_{S}$ are variables. Moreover, other empirical parameterizations have shown that $K_{M}$ depends upon Richardson number \cite[]{MG2010}.  While the inclusion of any relevant third parameter might improve the predictability of the classification scheme, the value of this improvement would have to be weighed against the added complexity of the resulting classification scheme.

\section{Conclusions} 

The equations governing the physics of estuarine circulation have been presented in non-dimensional form.  The two resulting non-dimensional parameters are the estuarine Froude number, $F_{R}$, 
and the modified tidal Froude number, $\widetilde{F_{T}}$.  Given these parameters the non-dimensional salinity gradient at the estuary mouth, $\bar{\Sigma}_{X}|_{0}$, and the non-dimensional salinity stratification (also at the estuary mouth), $\Phi_{0}$,
 can be computed.  The latter result forms the basis of a classification scheme that can be used to predict whether an estuary is fully or partially mixed, or highly stratified.  
The predictions of this classification scheme compare well with estuarine data.

\ifthenelse{\boolean{dc}}
{}
{\clearpage}
\bibliographystyle{ametsoc}
\bibliography{references}

\begin{thebibliography}{11}
\providecommand{\natexlab}[1]{#1}
\providecommand{\url}[1]{\texttt{#1}}
\providecommand{\urlprefix}{URL }
\expandafter\ifx\csname urlstyle\endcsname\relax
  \providecommand{\doi}[1]{doi:\discretionary{}{}{}#1}\else
  \providecommand{\doi}{doi:\discretionary{}{}{}\begingroup
  \urlstyle{rm}\Url}\fi
\providecommand{\eprint}[2][]{\url{#2}}

\bibitem[{Banas et~al.(2004)Banas, Hickey, MacCready, and Newton}]{Banas2004}
Banas, N., B.~Hickey, P.~MacCready, and J.~A. Newton, 2004: {Dynamics of
  Willapa Bay, Washington: a Highly Unsteady, Partially Mixed Estuary}.
  \textit{Journal of Physical Oceanography}, \textbf{34~(11)}, 2413--2427.

\bibitem[{Geyer(2010)}]{Estuarybook}
Geyer, W., 2010: {Estuarine Salinity Structure and Circulation}.
  \textit{{Contemporary Issues in Estuarine Physics}}, A.~Valle-Levinson, Ed.,
  Cambridge University Press, chap.~2, 12--26.

\bibitem[{Hansen and Rattray(1965)}]{HR1965}
Hansen, D.~V. and M.~J. Rattray, 1965: {Gravitational Circulation in Straits
  and Estuaries}. \textit{Journal of Marine Research}, \textbf{23~(2)},
  104--122.

\bibitem[{Hansen and Rattray(1966)}]{HR1966}
Hansen, D.~V. and M.~J. Rattray, 1966: {New Dimensions in Estuary
  Classification}. \textit{Limnology and Oceanography}, \textbf{11~(3)},
  319--326.

\bibitem[{Knudsen(1900)}]{Knud00}
Knudsen, M., 1900: {Ein hydrographischer Lehrsatz}. \textit{Annalen der
  Hydrographie und maritimen Meteorologie}, \textbf{28}, 316--320.

\bibitem[{MacCready(1999)}]{M1999}
MacCready, P., 1999: {Estuarine Adjustment to Changes in River Flow and Tidal
  Mixing}. \textit{Journal of Physical Oceanography}, \textbf{29~(4)},
  708--726.

\bibitem[{MacCready(2004)}]{M2004}
MacCready, P., 2004: {Toward a Unified Theory of Tidally-Averaged Estuarine
  Salinity Structure}. \textit{Estuaries}, \textbf{27~(4)}, 561--570.

\bibitem[{MacCready(2007)}]{M2007}
MacCready, P., 2007: {Estuarine Adjustment}. \textit{Journal of Physical
  Oceanography}, \textbf{37~(8)}, 2133--2145.

\bibitem[{MacCready and Geyer(2010)}]{MG2010}
MacCready, P. and W.~Geyer, 2010: {Advances in Estuarine Physics}.
  \textit{Annual Review of Marine Science}, \textbf{2~(1)}, 35--58.

\bibitem[{Prandle(1985)}]{Pran85}
Prandle, D., 1985: {On Salinity Regimes and the Vertical Structure of Residual
  Flows in Narrow Tidal Estuaries}. \textit{Estuarine, Coastal and Shelf
  Science}, \textbf{20~(5)}, 615--635.

\bibitem[{Ralston et~al.(2008)Ralston, Geyer, and Lerczak}]{Ralston2008}
Ralston, D., W.~Geyer, and J.~Lerczak, 2008: {Subtidal Salinity and Velocity in
  the Hudson River Estuary: Observations and Modeling}. \textit{Journal of
  Physical Oceanography}, \textbf{38~(4)}, 753--770.

\end{thebibliography}

\begin{table}[t]
\caption{Estimates of estuarine parameters calculated using the data of \cite{Pran85} and values of $B$ obtained from maps. Eq. (\ref{eq:Full2}) is used to obtain $\Phi_{0}$-theory.
}\label{tab:t1}
\begin{center}
\begin{tabular}{ccccccccc}
\hline\hline
 Estuary Name & $F_{R}$ & $F_{T}$ & $B/H$ &  $\widetilde{F_{T}}$ & $\Phi_{0}$  & $\Phi_{0}$-theory \\ \hline
Vellar  & $1.27$ & $ 0.64$ & $200$ & $9.0$ & $1.00$ &  $1.00$\\
Columbia  & $0.026$ & $ 0.43$ & $150$ & $ 5.3$ & $0.40$ &  $0.50$\\
James & $0.004$ &  $0.25$ & $360$ & $4.7$   & $0.22$ & $0.17$ \\
Tees  & $0.014$ & $1.03$ & $75$  & $8.9$ & $0.18$ & $0.33$\\
Southampton Waterway  & $0.0012$ & $0.37$ & $200$  &  $5.2$ & $0.10$ & $0.06$\\\
Tay & $0.014$ & $1.38$ & $400$  & $27$  & $0.10$ & $0.17$\\
Narrows of the Mersey   & $0.0009$ & $0.83$ & $65$  & $6.7$ & $0.05$ & $0.05$\\
Bristol Channel  & $0.006$ & $1.59$ & $300$ & $27$ & $0.02$ & $0.06$\\
\hline
\end{tabular}
\end{center}
\end{table}

\begin{table}[t]
\caption{List of coefficients used in different equations.
}\label{tab:t2}
\begin{center}
\begin{tabular}{ccccccccc}
\hline\hline
 Coefficient & Value \\ \hline
 $C_{1}$ & $1.67Sc^{1/3}$ \\
 $C_{2}$ & $0.792Sc^{2/3}$\\
 $C_{3}$ &  $41.7a_{0}a_{1} C_{D} Sc^{-1/3}$\\
 $C_{4}$ & $0.868Sc^{-1/3}$ \\
 $C_{5}$ & $36.2Sc^{1/3}$\\
 $C_{6}$ & $41.7Sc^{2/3}$ \\
 $C_{7}$ & $4.08Sc^{1/3}$\\
 $C_{8}$ & $3.57Sc^{2/3}$ \\
 $C_{9}$ & $5.43Sc^{1/3}$ \\
 $C_{10}$ & $5.21Sc^{2/3}$ \\
\hline
\end{tabular}
\end{center}
\end{table}

\end{document}